\documentclass[10pt,a4paper,english,journal]{IEEEtran}
\usepackage[T1]{fontenc}
\usepackage[latin9]{inputenc}
\usepackage{amstext}
\usepackage{graphicx}

\makeatletter

\pdfpageheight\paperheight
\pdfpagewidth\paperwidth

\providecommand{\tabularnewline}{\\}

\usepackage{xcolor,soul}
\sethlcolor{lightgray}

\makeatother

\usepackage{babel}
\begin{document}
\title{Capacity Cost of Fulfilling the URLLC Performance in Industrial 5G New Radio Deployments}
\author{\IEEEauthorblockN{Ali A. Esswie\textit{ }\\
Cellullar Standards Technical Lead, Advanced Air Interface, Future Wireless, InterDigital Communications.
ali.esswie@interdigital.com}}

\maketitle
$\pagenumbering{gobble}$
\begin{abstract}
The development of the 5G new radio specifications has been derived
by the the deterministic low latency use-cases such as the ultra-reliable
and low-latency communications (URLLC). A URLLC application requires a stringent
radio latency and reliability performance, e.g., one-way radio latency
of 1 ms with 99.999\% success probability. Furthermore, there is a
concurrent progressive demand for broadband capacity cellular applications,
e.g., enhanced mobile broadband (eMBB) use-cases. The coexistence
among the URLLC and eMBB service classes over a single radio spectrum
is a challenging task since achieving the tight URLLC radio targets
typically results in a capacity loss. Hence, it is vital for telecom
operators to understand the capacity cost of fulling the various URLLC
requirements in order to sufficiently plan the corresponding pricing
models. Hence, in this work, a comprehensive analysis of the system
capacity loss is presented to achieve the various requirements of the different URLLC use-cases.
An extensive set of realistic system level simulations is performed
and introduced where valuable insights and system design recommendations
on the URLLC-eMBB quality of service coexistence are presented.  

\end{abstract}

\section{Introduction}

Unlike the former radio network generations, the fifth generation
(5G) new radio (NR) specifications support revolutionary multi quality-of-service
(QoS) classes such as the ultra-reliable low-latency communications (URLLC),
and the enhanced mobile broadband (eMBB) {[}1{]}. The URLLC applications
demand a stringent set of the radio latency and reliability targets while
the eMBB services demand broadband communication data rates {[}2{]}.
This multi-QoS-class coexistence enables novel cellular applications
and use-cases, e.g., tactile internet, true smart infrastructures,
and virtual reality communications {[}3{]}. 

However, the deterministic low-latency industrial applications, i.e.,
industry 4.0 deployments {[}4, 5{]}, are the lead drivers of the 5G
and 5G-advanced (release-18 and beyond) developments. Within those
deployments, a wide range of; though, conflicting, QoS classes, radio
performance requirements, and device types should be supported. That
is, a deterministic ultra-low radio latency with a certain reliability
level is always guaranteed while a pre-determined capacity/data rate
target(s) should be preserved for simultaneous capacity demanding
services {[}6{]}. Fundamentally, achieving an ultra-low radio latency
leads to an insufficient radio capacity of the same communication
bandwidth. Thus, it is of a significant importance for telecom operators
and cellular service providers to identify the system capacity cost,
which is paid/lost in order to fulfill a certain ultra-low latency
service performance target, and hence, adapt their pricing models
accordingly. 

In the state-of-the-art open literature, there are many contributions
of MAC and PHY schemes that trade-off the achievable system capacity
with the lowest guaranteed radio latency. Examples include reinforcement
learning based link adaptation for a faster URLLC packet transmission
{[}7, 8{]}, QoS-aware scheduling schemes {[}9{]}, and on-the-fly resource
preemption techniques {[}1, 10{]}. Those proposals seek to achieve
a guaranteed maximum radio latency for the URLLC critical services
while maximizing the overall system capacity, i.e., by incurring the
minimum possible capacity loss of the eMBB traffic. 

In this paper, a comprehensive analysis of the capacity loss, due
to fulfilling a certain URLLC latency and reliability performance
target, is presented. Various URLLC latency and reliability targets
are considered to match the different URLLC use-cases and deployments.
The reference case considered in this work is the best effort eMBB, where the corresponding radio latency performance is relaxed.
Thus, the paper answers the following question: ''what is the system
capacity lost, compared to the best effort case, to satisfy a guaranteed
maximum URLLC radio latency with a certain radio reliability level?''.
An extensive set of realistic system level simulations is performed
to obtain a set of statistically reliable results and the respective conclusions.

This paper is organized as follows. Section II presents the system
model and the main performance indicator considered in this work.
Section III introduces the simulation methdology and the performance
results. Section IV presents the acknowledgments and Section IV concludes the paper.

\section{System Model}

\subsection{Setting the scene}

In this work, we consider an industrial factory automation deployment
with $C$ indoor cells, horizontally inter-distanced by $d=20$ meters,
as shown by Fig. 1. Each cell serves an average number of $K$ uniformly
distributed user-equipment's (UEs), where even UE dropping for all
cells is adopted. In this study, we assume only downlink (DL) traffic
towards active UEs. To emulate the URLLC use cases, the traffic is
characterized by the FTP3 traffic model, where data packets of a finite
payload size $\textnormal{B-Bytes}$ are considered. The DL packets
arrive at the cell following a Poisson Arrival Process with a predefined
mean arrival rate $\lambda$ (packets/sec). Therefore, the total offered
load $\Omega$ in bits/sec for the entire factory deployment is calculated
as:

\begin{figure}
\begin{centering}
\includegraphics[scale=0.9]{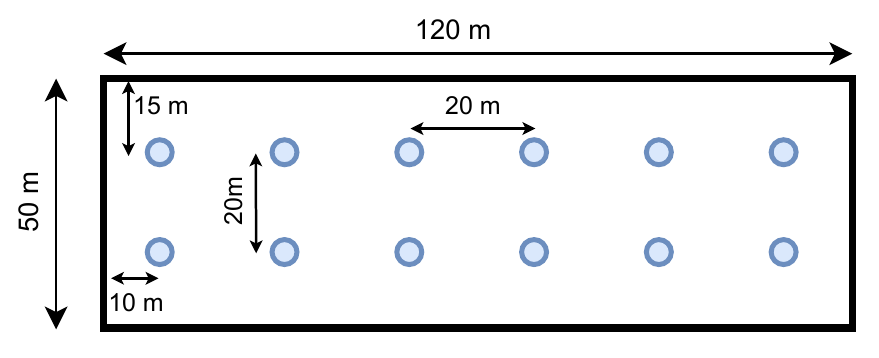}
\par\end{centering}
\begin{centering}
\caption{Industrial factory deployment.}
\par\end{centering}
\end{figure}

\begin{equation}
\Omega=C\times\left(K\times B\times8\times\lambda\right).
\end{equation}

We follow the 3GPP assumptions and guidelines for URLLC simulations.
URLLC UEs are dynamically multiplexed using the orthogonal frequency
division multiplexing (OFDMA). The 30 kHz numerology (i.e., sub-carrier
spacing) is used in this work, in line with {[}1{]}, since it offers
a sufficiently short OFDM symbol duration in order to fulfill the
stringent radio latency requirements of the URLLC services. The minimum
schedulable resource unit is the physical resource block (PRB), consisting
of 12 successive sub-carries. Moreover, we consider a short transmission
time interval (TTI) duration of 4 OFDM symbols for faster URLLC transmission
and scheduling. 

Thus, an arriving DL packet is processed as follows: the packet is
first received and prepared for transmission by the serving cell RF
stack processor. The delay consumed before the DL transport block
is ready for transmission is taken explicitly into account in this
study, in line with {[}2{]}. Then, the dynamic scheduler of the serving
cell multiplexes all pending DL packets for transmission, using the
proportional fair (PF) scheduling criterion. Hence, packets of the
highest PF metric, are transmitted during the current TTI interval
while other packets are further buffered towards the upcoming scheduling
instants, i.e., queuing delay of the dynamic scheduling. We consider
an adaptive modulation and coding selection (MCS) corresponding to
a first-transmission block error rate (BLER) of 1\%. Therefore, the
MCS level is dynamically selected based on the latest reported channel
quality indication (CQI) from each UE. 

Accordingly, the serving cell sends a scheduling grant, using the
lower-layer downlink control information (DCI) signaling, to notify
the corresponding UEs of the DL resource allocations. The resource
overhead of the physical-layer control signaling is explicitly taken
into account; however, in this work, we assume the delay of transmitting
and processing the scheduling grants is negligible. At the UE side,
the delay consumed to process and decode received DL packets is explicitly
considered into account of the total radio latency. If the transmitted
DL packet is not successfully decoded at the UE, the UE sends a hybrid
automatic repeat request (HARQ) negative acknowledgment (NACK). The
delay for transmitting the HARQ feedback for each packet is also explicitly
considered. In such case, the serving cell re-transmits the respective
packet for soft-combining at the UE side. At the cell side, the HARQ
re-transmissions are always prioritized over new packet transmission
in order to reduce the overall radio latency for pending packets.

\subsection{Main performance metric indicators (KPIs)}

The major KPIs of this work are the achievable one-way URLLC outage
latency and the corresponding network throughput/capacity, respectively. Thus, the
total radio latency of each DL packet $\varphi$ is measured and tracked,
for various outage probabilities $\rho$. In particular, the considered
latency metric denotes the delay from the moment a DL packet is generated
and arrives at the packet data convergence protocol (PDCP) layer of
the serving cell until it is successfully decoded at the intended
UE. This sums up the cell and UE processing delays, the transmission
delay, the dynamic scheduling queuing delay, and HARQ re-transmission
delay, respectively. Therefore, for a certain target outage probability,
we present the maximum supported offered load/capacity of the network
such as to fulfill a maximum guaranteed URLLC radio latency. 

Furthermore, in this work, we adopt the throughput cost metric $\mathit{\Psi}$
of the URLLC and best effort (BE) cases, respectively. The BE case
denotes the transmissions of an infinite packet size and without a
target outage latency requirement such that the network capacity is
maximized. Hence, the cost metric implies how much network throughput
is lost (paid) in order to fulfill a certain URLLC outage latency
target, compared to the achievable network throughput of the BE case,
and is calculated as follows:

\begin{equation}
\mathit{\Psi}\left(B,\varphi,\rho\right)=\left(1-\frac{\mu_{urllc}\left(B,\varphi,\rho\right)}{\mu_{BE}}\right)\times100\%,
\end{equation}
where $\mathit{\Psi}\left(B,\varphi,\rho\right)$ denotes the inflicted
network throughput cost of the URLLC use cases, with a packet size
of $\textnormal{B-Bytes}$ in order to fulfill a maximum guaranteed
radio latency of $\varphi$ ms for the target outage probability $\rho$.
$\mu_{urllc}$ and $\mu_{BE}$ are the achievable mean network throughput
metrics of the URLLC and BE cases, respectively. 

\section{Performance Evaluation}

\subsection{Simulation Methodology}

We comprehensively evaluate the performance of the URLLC service class using extensive system level simulations. The major system settings are presented in Table I. The conducted simulations follow the system modeling assumptions, presented in Section II, and the general 3GPP simulation methodology. The industrial factory deployment layout is considered alongside with employing the state of the art industrial factory channel model. Dynamic FTP3 traffic arrivals are considered at each UE, where the arrival rate follows a Poisson Point Arrival Process. When a UE is created, it connects to the surrounding cell with the highest received reference signal power (RSRP). The simulations explicitly include the major functionalities of the PHY and MAC stack layers. For each transmitted packet, the sub-carrier signal to interference noise ratio (SINR) is calculated using the linear minimum mean squared error interference rejection and combining (LMMSE-IRC) receiver. The effective SINR is calculated by combining the estimated sub-carrier SINRs using the mean mutual information per coded bit (MMIB) mapping. Based on the effective SINR, the packet error probability (PEP) is calculated from predefined look-up tables, obtained from extensive link level simulations. Therefore, based on PEP, the packet is determined as successfully received or not. If the packet is not successfully decoded from first transmission, the corresponding HARQ re-transmissions are triggered. Every DL TTI, UEs are dynamically scheduled based on the proportional fair criterion. The corresponding modulation and coding scheme (MCS) is selected based on the latest available channel quality indicator (CQI) reports. 

Finally, the simulator is periodically calibrated by reporting and comparing baseline performance statistics among the 3GPP partners. Moreover, in line with [2], we run simulations for a sufficiently long period of time in order to ensure high statistical confidence of the achievable results before drawing solid conclusions. That is, the default simulation time is at least 5 million successfully-decoded URLLC packets.

\begin{table}
\caption{{\small{}Simulation parameters.}}

\centering{}%
\begin{tabular}{c|c}
\hline 
Parameter & Value\tabularnewline
\hline 
Environment & 3GPP-InF, one cluster, 12 cells\tabularnewline
\hline 
DL channel bandwidth & 40 MHz, SCS = 30 KHz, FDD\tabularnewline
\hline 
Channel model & InF-DH (dense clutter and high BS) {[}11{]}\tabularnewline
\hline 
BS transmit power & BS: 25 dBm\tabularnewline
\hline 
Carrier frequency & 4 GHz\tabularnewline
\hline 
BS  heights & BS: 10m\tabularnewline
\hline 
Antenna setup & $2 x 2$ \tabularnewline
\hline 
Average UEs per cell &   10\tabularnewline
\hline 
TTI configuration & 4-OFDM symbols\tabularnewline
\hline 
URLLC Traffic model & \textnormal{FTP3,  paket size = 50/1500 Bytes}\tabularnewline
\hline 
eMBB Traffic model & Best effort with infinite payload size\tabularnewline
\hline 
DL scheduling & proportional fair\tabularnewline
\hline 
Processing time & $\begin{array}{c}
\textnormal{PDSCH prep. delay: 2.5-OFDM symbols}\\
\textnormal{PDSCH decoding : 4.5-OFDM symbols}\\
\end{array}$\tabularnewline
\hline 
DL receiver & 	L-MMSE-IRC\tabularnewline
\hline 
\end{tabular}
\end{table}

\subsection{Performance Results}

In this section, we mainly evaluate the inflicted network capacity
loss in order to fulfill certain URLLC outage latency targets. In
particular, the maximum supported offered capacity is investigated
for URLLC and best effort cases, respectively. The URLLC services
are evaluated to achieve various outage latency targets. That is,
for different maximum guaranteed radio latency budgets for several
outage probability levels, and using two payload sizes (50 and 1500
Bytes). The best effort reference denotes the case where the requirement
on the radio latency is relaxed, and is evaluated under two different
UE schedulers: the proportional fair (PF) and equal throughput (ET)
schedulers, respectively. The former enables a fair scheduling criterion
without an inter-UE throughout regularization while the latter seeks
to achieve a guaranteed equal throughput per UE. 

Fig. 2 depicts the maximum supported offered load of the URLLC traffic,
with a large 1500-byte payload size, to achieve a maximum guaranteed
outage latency of 1, 3 and 10 ms, respectively, and under various
outage probabilities as $10^{-5}$ and $10^{-2}$. As can be noticed,
achieving the 1 ms latency deadline with $10^{-5}$ outage probability
significantly degrades the achievable capacity compared to both cases
of the best effort class, with PF and ET schedulers, respectively.
That is, to achieve a maximum guaranteed 1 ms of the radio latency,
the maximum supported offered loads are 2.03 Mbps and 11.65 Mbps for
the $10^{-5}$ and $10^{-2}$ outage probabilities, compared to 71.06
Mbps and 93 Mbps with the best effort case of the ET and PF packet
schedulers. Furthermore, the best effort case with ET scheduler clearly
exhibits \textasciitilde{} 23.6\% reduction of the achievable capacity
compared to the case with the PF scheduler, due to sacrificing part
of the system capacity in order to guarantee an equal-UE perceived
throughput. 

\begin{figure}
\begin{centering}
\includegraphics[scale=0.55]{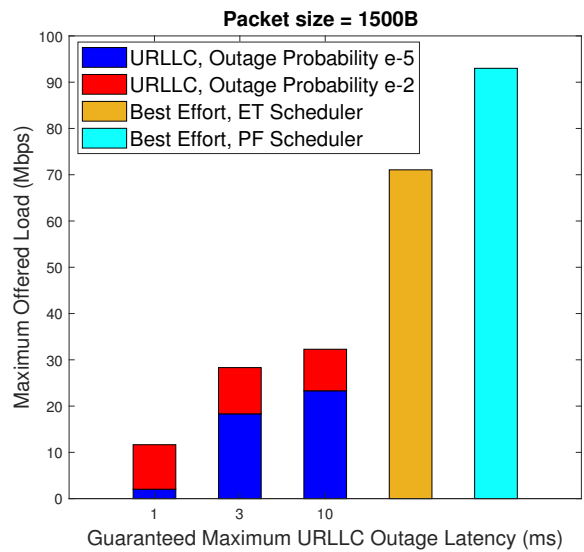}
\par\end{centering}
\caption{URLLC outage latency performance, large payload case.}

\end{figure}

Looking at the case of the URLLC small payload, Fig. 3 shows the achievable
load performance of the 50-byte URLLC case, where similar conclusions
as of Fig. 1 are observed. Relaxing the stringent requirement of the
outage probability offers a clear capacity improvement over the low
radio latency region. For instance, to achieve a maximum guaranteed
1-ms target, the supported offered capacity is improved by \textasciitilde{}
76.4\% with the $10^{-2}$ outage probability, compared to that is
of the $10^{-5}$ outage level. 

\begin{figure}
\begin{centering}
\includegraphics[scale=0.55]{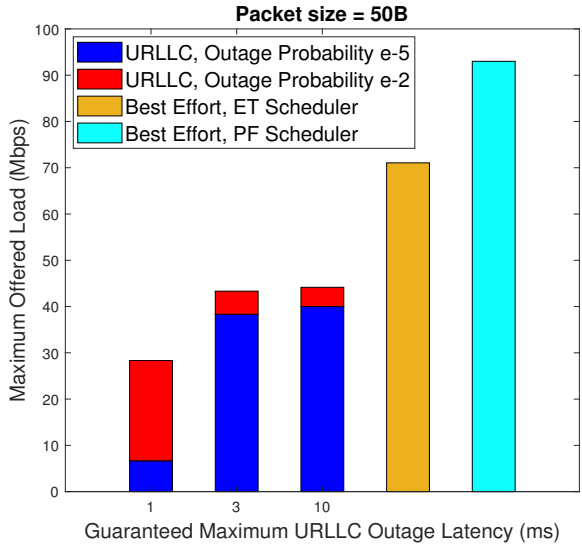}
\par\end{centering}
\caption{URLLC outage latency performance, small payload case.}
\end{figure}

The packet size is deemed to have a vital impact on the achievable
joint outage latency and capacity performance. Therefore, Fig. 4 depicts
a capacity comparison of the URLLC class with small and large payload
sizes, i.e., 50 and 1500 bytes, respectively, at the $10^{-5}$ outage
probability. As clearly seen, the small packet transmissions are observed
to be more efficient in supporting more offered capacity to achieve
the same outage latency target, compared to that is of the large payload.
For instance, to achieve the stringent 1-ms latency target, the small-payload
transmissions supports \textasciitilde{} 56.4\% more offered load
than the large-payload case. 

\begin{figure}
\begin{centering}
\includegraphics[scale=0.55]{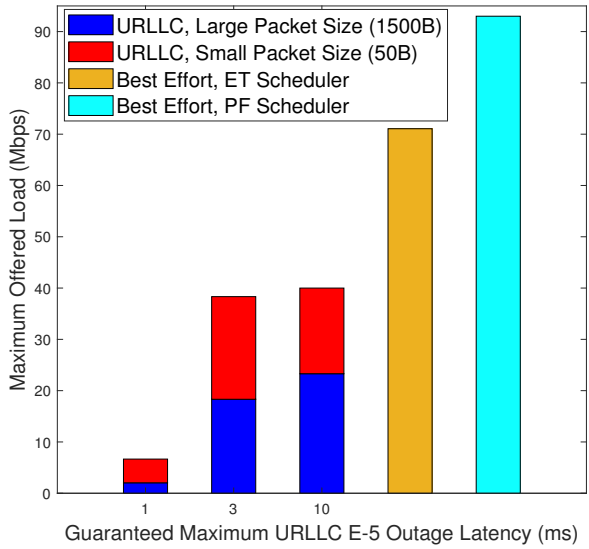}
\par\end{centering}
\caption{URLLC outage latency performance, with payload size.}
\end{figure}

This is mainly attributed to the required multiple resource allocations
over multiple TTIs for the large-size packets to get fully transmitted.
Fig. 5 depicts the empirical cumulative distribution function (ECDF)
of the total number of scheduled PRBs per each FTP3 packet, for different
supported offered load levels, which correspond to a certain maximum
guaranteed outage latency, i.e., 1, 1.5, and 3 ms radio latency targets,
respectively, at the $10^{-5}$ outage probability. At the 50\%ile
level, a single 1500-byte URLLC packet requires 63 and 78 PRBs (out
of total 100 PRBs) of a single TTI, respectively, for the three latency
targets under evaluation to be fulfilled. This implies a maximum of
a single packet transmission per TTI while the other concurrent packets
from other active UE are being buffered towards the upcoming transmission
opportunities. However, at the tail distribution, i.e., 95\%ile level,
for the offered loads of 70 and 220 Mbps, respectively, a single FTP3
packet requires 138 and 237 PRBs, respectively. This denotes that
a single 1500-byte packet is scheduled over multiple TTIs, consuming
most of the available bandwidth. This is particularity relevant to
the packets from the cell-edge UEs, where their decoding ability is
highly deteriorated due to the strong inter-cell interference. Thus,
to achieve the target 1\% BLER, the serving BS relies on a conservative
MCS selection for the respective packet transmissions, leading to
a significant degradation of the network spectral efficiency. 

\begin{figure}
\begin{centering}
\includegraphics[scale=0.49]{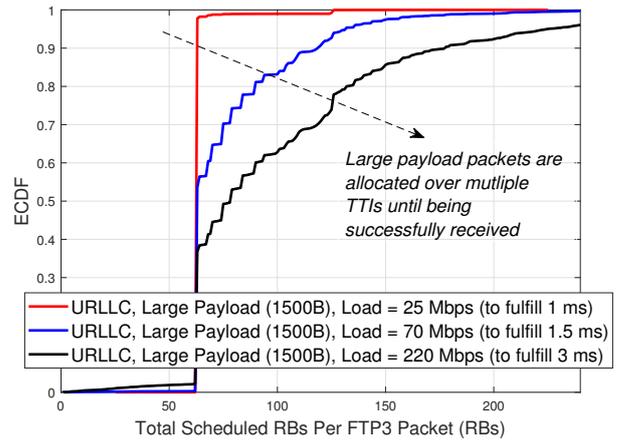}
\par\end{centering}
\caption{Resource utilization performance of the URLLC large packets, for different
outage latency targets.}
\end{figure}

On the opposite side, and as shown by Fig. 6, the small-payload URLLC
transmissions, i.e., 50-byte payload, are more efficient than the
large 1500-byte payload from the minimum-delay allocation perspective.
To fulfill the stringent 1-ms latency target at the $10^{-5}$ outage
probability, the small-payload transmissions require 3 PRBs on average
compared to 63 PRBs with the large payload. This allows for: (a) scheduling
the entire small-payload URLLC packets within a single TTI duration
without segmentation, and (b) co-scheduling multiple packets from
multiple active UEs at the same TTI, hence, reducing the average packet
queuing delay accordingly. 

\begin{figure}
\begin{centering}
\includegraphics[scale=0.49]{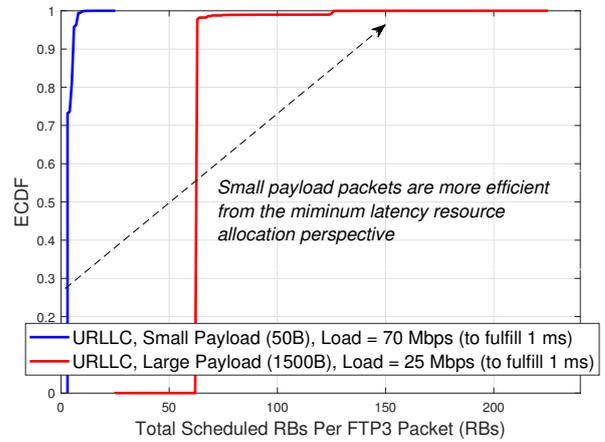}
\par\end{centering}
\caption{URLLC resource utilization performance, for small and large payload
packets.}
\end{figure}

Finally, Fig. 7 and 8 show the overall throughput cost of the URLLC
service class, compared to the best effort case. We consider various
outage levels and latency budgets of the large and small payload URLLC
cases, respectively. In particular, such cost denotes how much network
throughput is lost (paid), relative to the best effort case, in order
to fulfill the corresponding stringent URLLC outage latency target.
For example, with the large-payload URLLC case in Fig. 7, the throughput
cost spans the range from 20\% up to 97\%. To fulfill a maximum guarantee
latency of 0.5 ms while employing the URLLC large payload, the supported
network offered load is 97\% less than that is of the best effort
case. This is majorly because, with the large payload URLLC transmissions,
and to fulfill such stringent latency deadline, URLLC packets must
be scheduled within a single TTI duration, and without further packet
queuing or re-transmissions. Therefore, a single large-payload URLLC packet scheduling
consumes the entire bandwidth, i.e., the majority of the available
PRBs. Hence, the maximum supported offered capacity is drastically
reduced compared to the best effort case. Similar conclusions can be drawn from Fig. 8, for the case of the URLLC small payload size, i.e., 50 Bytes.

\begin{figure}
\begin{centering}
\includegraphics[scale=0.5]{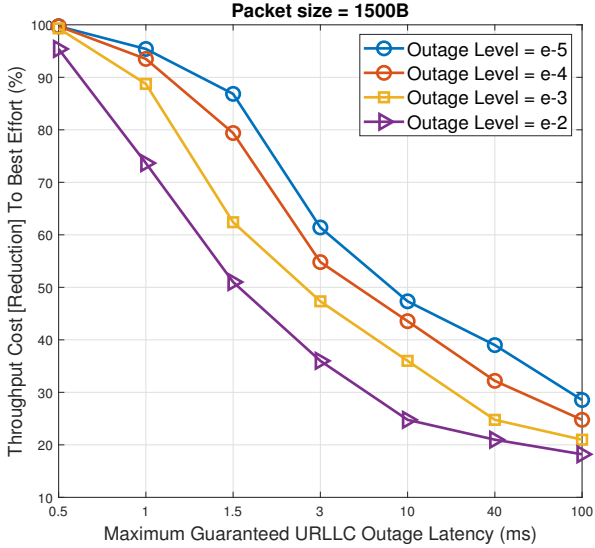}
\par\end{centering}
\caption{Spectral efficiency cost of achieving the URLLC outage targets, large
payload size.}
\end{figure}

\begin{figure}
\begin{centering}
\includegraphics[scale=0.5]{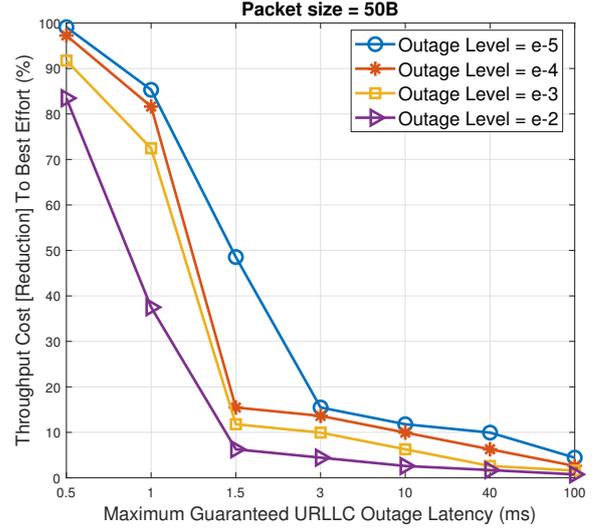}
\par\end{centering}
\caption{Spectral efficiency cost of achieving the URLLC outage targets, small
payload size.}
\end{figure}

\section{Acknowledgments}

This work has been partially completed while the main author was with Nokia Bell Labs, Standardization and Research Access Lab, Aalborg, Denmark.

\section{Concluding Remarks }

In this paper, a comprehensive analysis of the system capacity loss,
due to satisfying the various URLLC targets, is presented by means
of extensive system level simulations. The paper answers the following
research question: ''what is the network capacity loss, compared to
best effort case, due to fulfilling a guaranteed maximum URLLC latency
and reliability target?'', and hence, the paper offers valuable insights
for telecom operators to adapt their pricing models within their multi-QoS
5G deployments. Various URLLC configurations and settings are considered
to match the different realistic URLLC use cases, and the achievable
system capacity is compared to that is of the best-effort case, where
the radio latency performance is relaxed.


\begin{thebibliography}{10}
\bibitem[1]{key-1} A. A. Esswie and K. I. Pedersen, \textquotedbl Opportunistic
Spatial Preemptive Scheduling for URLLC and eMBB Coexistence in Multi-User
5G Networks,\textquotedbl{} in IEEE Access, vol. 6, pp. 38451-38463,
2018.

\bibitem[2]{key-2} A. A. Esswie and K. I. Pedersen, \textquotedbl On
the Ultra-Reliable and Low-Latency Communications in Flexible TDD/FDD
5G Networks,\textquotedbl{} \emph{in Proc. IEEE CCNC}, 2020, pp. 1-6.

\bibitem[3]{key-3} A. Ghosh, A. Maeder, M. Baker and D. Chandramouli,
\textquotedbl 5G Evolution: A View on 5G Cellular Technology Beyond
3GPP Release 15,\textquotedbl{} in IEEE Access, vol. 7, pp. 127639-127651,
2019.

\bibitem[4]{key-4} M. Gundall et al., \textquotedbl Introduction
of a 5G-Enabled Architecture for the Realization of Industry 4.0 Use
Cases,\textquotedbl{} in IEEE Access, vol. 9, pp. 25508-25521, 2021.

\bibitem[5]{key-5} W. Trneberg et al., \textquotedbl Towards Intelligent
Industry 4.0 5G Networks: A First Throughput and QoE Measurement Campaign,\textquotedbl{}
\emph{in Proc. IEEE SoftCOM}, 2020, pp. 1-6.

\bibitem[6]{key-6} A. A. Esswie and K. I. Pedersen, \textquotedbl Analysis
of Outage Latency and Throughput Performance in Industrial Factory
5G TDD Deployments,\textquotedbl{} \emph{in Proc. IEEE VTC-Spring}, 2021,
pp. 1-6.

\bibitem[7]{key-7} A. A. Esswie, K. I. Pedersen and P. E. Mogensen,
\textquotedbl Online Radio Pattern Optimization Based on Dual Reinforcement-Learning
Approach for 5G URLLC Networks,\textquotedbl{} in IEEE Access, vol.
8, pp. 132922-132936, 2020.

\bibitem[8]{key-8} C. She et al., \textquotedbl A Tutorial on Ultrareliable
and Low-Latency Communications in 6G: Integrating Domain Knowledge
Into Deep Learning,\textquotedbl{} in Proceedings of the IEEE, vol.
109, no. 3, pp. 204-246, March 2021.

\bibitem[9]{key-9} L. Wang and H. Zhang, \textquotedbl Analysis
of joint scheduling and power control for predictable URLLC in industrial
wireless networks,\textquotedbl{} \emph{in Proc. IEEE ICII}, Orlando, FL,
USA, 2019, pp. 160-169

\bibitem[10]{key-10} M. Morcos, M. Mhedhbi, A. Galindo-Serrano and
S. Eddine Elayoubi, \textquotedbl Optimal resource preemption for
aperiodic URLLC traffic in 5G Networks,\textquotedbl{} \emph{in Proc. IEEE
PIMRC}, 2020, pp. 1-6.

\bibitem[11]{key-11} \emph{Study on channel model for frequencies from 0.5 to 100 GHz}; Release
16, 3GPP, TR 38.901, V16.1.0, Dec. 2019.
\end{thebibliography}
\end{document}